\documentclass[conference]{IEEEtran}

\usepackage[latin1]{inputenc}
\usepackage[english]{babel}
\usepackage{amsmath}
\usepackage{amsfonts}
\usepackage{amssymb}
\usepackage[lofdepth,lotdepth]{subfig}
\usepackage{makeidx}
\usepackage{graphicx}
\usepackage{charter}
\usepackage{algorithmic}
\usepackage{algorithm}
\usepackage{listings}
\usepackage{pdflscape}
\usepackage{tabularx}
\usepackage[colorlinks=false]{hyperref}
\usepackage{cite}
\usepackage{pifont}
\usepackage{multicol}
\usepackage{amsfonts}

\newcommand{\comment}[1]{}

\begin{document}
\title{Finding Model-Checkable Needles in Large Source Code Haystacks: Modular Bug-Finding via Static Analysis and Dynamic Invariant Discovery}

\author{\IEEEauthorblockN{%
Mohammad Amin Alipour,\IEEEauthorrefmark{1} Alex
Groce,\IEEEauthorrefmark{1} Chaoqiang Zhang,\IEEEauthorrefmark{1}
Anahita Sanadaji,\IEEEauthorrefmark{1} and Gokul
Caushik,\IEEEauthorrefmark{1} }

\IEEEauthorblockA{\IEEEauthorrefmark{1}%
School of Electrical Engineering and Computer Science\\
Oregon State University\\
\{alipourm,grocea,zhangch,alipourm,sanandaa,caushikg\}@onid.oregonstate.edu%
}
}

\maketitle

\begin{abstract}
In this paper, we present a novel marriage of static and dynamic
analysis.  Given a large code base with many functions and a mature
test suite, we propose using static analysis to find functions 1) with
assertions or other evident correctness properties (e.g., array bounds
requirements or pointer access) and 2) with simple enough control flow
and data use to be amenable to predicate-abstraction based or bounded
model checking without human intervention.  Because most such
functions in realistic software systems in fact rely on many input
preconditions not specified by the language's type system (or
annotated in any way), we propose using dynamically discovered
invariants based on a program's test suite to characterize likely
preconditions, in order to reduce the problem of false positives.
While providing little in the way of verification, this approach may
provide an additional quick and highly scalable bug-finding method for
programs that are usually considered ``too large to model check.''  We
present a simple example showing that the technique can be useful for
a more typically ``model-checkable'' code base, even in the presence
of a poorly designed test suite and bad invariants.
\end{abstract}

While new tools and techniques are being devised to conquer software
verification problems, software systems themselves are becoming more
complicated, with multiple layers of software/hardware interactions,
functioning in dynamic, uncertain environments.  Even where teams of
model checking experts are in place, and formal verification is
acknowledged as a valuable practice, model checking is seldom applied
to complete modern code bases for realistic-scaled systems.  For
example, while model checking was applied to small portions (typically
10,000 lines or less) of the code base for NASA/JPL's Curiosity
rover~\cite{CFV08}, even the Laboratory for Reliable Software did not
attempt to apply model checking to the full approximately 2 million
lines of Curiosity software.

\begin{figure*}[t]
\centering
\includegraphics[width=4.0in]{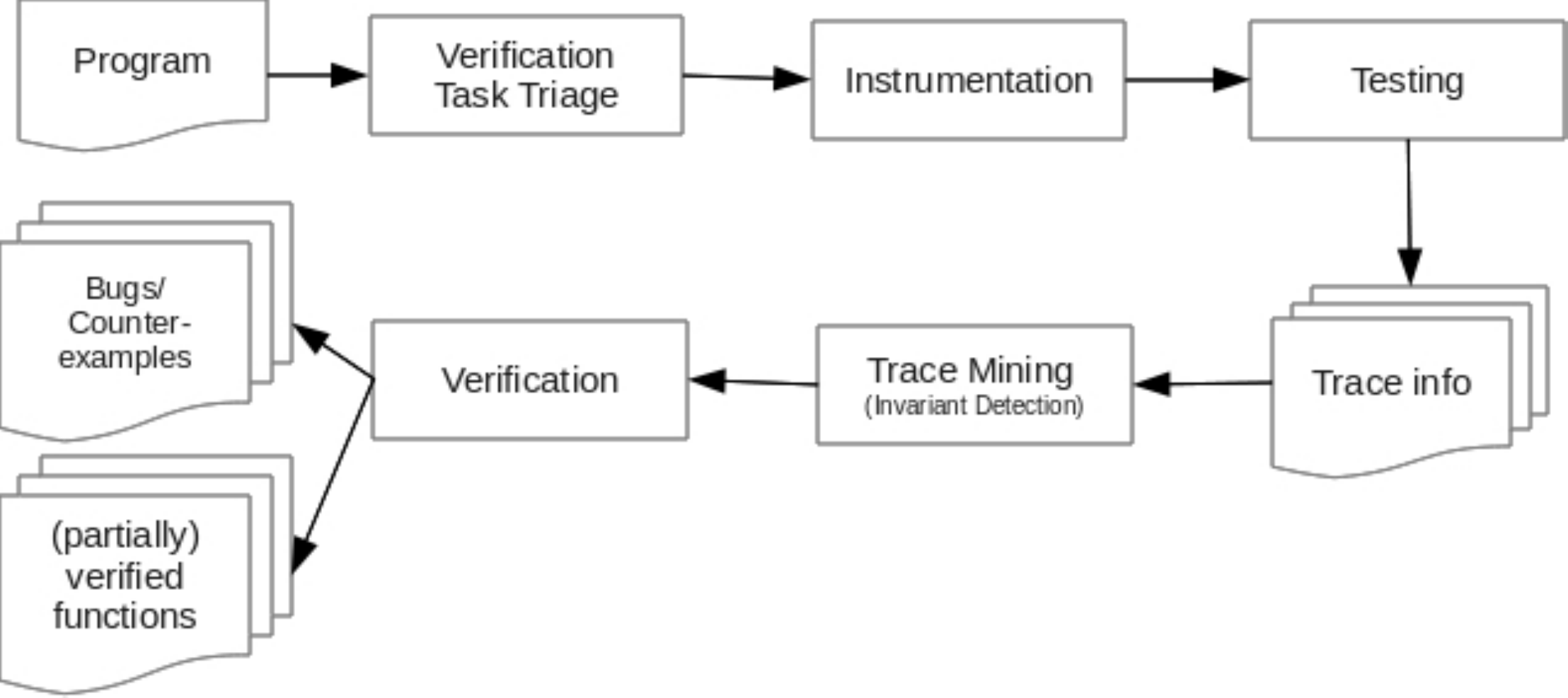}
\caption{The core workflow of our modular verification approach}
\label{fig:idea}
\end{figure*}

Static analysis methods, on the other hand, are expected to scale to
real-world code bases.  Even the popular technical
press~\cite{CoverityCuriosity} has noted the use of Coverity and other
tools on huge code bases for the Curiosity rover, the CERN particle
accelerator, and other big mission-critical systems.

While model checking can be applied to whole systems when a research
team drives the system design and real-world deadlines and constraints
are removed, its role in most critical system development efforts is
typically relegated to verifying (or finding bugs in) small
components.  One approach to escaping this limitation is to enhance
the fundamental scalability of model checking by using abstraction or
other techniques to manage huge state spaces, and especially by using
compositional
reasoning~\cite{LearnReuse,cobleigh2006breaking,gheorghiu2008automated,MAGIC}
to handle large code bases.  Such efforts are an important
contribution, but thus far have yet to crack anything like the 1
million LOC mark in an automated fashion.

Beyond the fundamental problem of the state space explosion, there is
a second scalability problem with most current model checking tools:
they are far less ``industrially hardened'' than commercial static
analysis tools, and fail for a variety of essentially engineering
reasons for very large code bases --- partly because they are seldom
tested on more than a few thousand lines of code at a time.  Model
checking tools work best, and most automatically, when given code
fragments only the size of a few small functions, in which case both
predicate abstraction~\cite{MAGIC} and bounded model
checking~\cite{BMC,cbmc} can be used even by upper-level undergraduate
students to verify properties and find subtle bugs.  This fact may lie
behind the success of model checkers for finding bugs in container
classes~\cite{Willemissta2006}.

This paper proposes and presents very preliminary results for a novel
``modular'' method for applying model checking automatically to
real-world code bases that does not require conquering the state-space
explosion problem, or automatically decomposing such a system into
components in the traditional fashion for verification.  We propose to
exploit precisely the current strengths of model checking by
\emph{mining a large code base for small, self-contained, portions of
code suitable for push-button model checking}.  The core idea is to
use static analysis to identify functions (or small sets of functions)
without recursion or recursive datatypes, overly complex pointer
manipulations, or whatever other features typically frustrate model
checking efforts.  Additionally, such code fragments should contain
assertions or other statements (e.g. array or pointer accesses, system
calls) that make it easy to generate properties to check without human
intervention.  We call our search \emph{verification task triage}: our
contribution is to propose that one effective way to apply model
checking to large systems is to simply cherry pick the parts of the
system that \emph{can be model checked.}  This turns on its head the
usual effort to apply human effort to verifying the components of most
interest in a large system, which can proceed independently.  Our
approach to ``modular verification'' works more like many static
analysis algorithms, which ``give up'' on pointers when alias analysis
fails, and abandon overly complex paths.  A more fanciful way to
understand the idea is to consider the not infrequent occasion of a
graduate student or model checking researcher combing a large,
critical system (e.g. OpenSSL) for some functions suitable for
specification and verification.  We aim to replace such a human with a
static analysis algorithm that is less intelligent, but much faster
and able to generate many likely candidates rather than being
satisfied after finding one or two functions.

Unfortunately, naive verification task triage plus model checking is unlikely to
yield effective results, even for the purpose of bug finding.  We
believe that most functions in complex code bases rely on
preconditions over inputs and data values accessed that are not
represented by the language's type system or by other annotations.
Model checking without these preconditions will probably lead to an
unmanageable number of uninteresting bug reports --- the false
positives that have always plagued static analysis efforts.

We propose to mitigate this problem by using dynamic invariant
detection~\cite{daikon} in a (to our knowledge) novel way: as a tool
for generating approximations of pre-conditions needed to avoid false
positives in large code bases.  The idea, shown in Figure
~\ref{fig:idea}, involves a two-stage workflow.  First, we perform
verification task triage to identify suitable small pieces of code to
model checking a large code base.  Then, the test suite for the large
system is run, with instrumentation inserted to extract trace
information over the suitable functions.  Dynamic invariant generation
takes program traces and identifies, via ``empirical'' rather than
static means, invariants --- propositions that are true for all
observed executions of the system.  These invariants are then used to
annotate the triage-identified functions with ``preconditions''
generalizing the data values seen during execution of the test suite.
The annotated pieces of code are model checked, with the expectation
that any bugs detected are likely to be real faults since the
invariants ensure that data ranges and relationships are like those
observed over the test suite.  As a simple example, consider the case
of a function {\tt f} that takes as one input a value {\tt int m}.  In
typical world-world C code, {\tt m} may in fact represent a highly
constrained quantity, such as the number of minutes since midnight,
but the type system will not express this constraint.  Model checking
{\tt f} without this information may produce a counterexample
involving, for example, integer overflow due to multiplying {\tt m} by
60.  Even if a real off-by-one bug exists when {\tt m = 1}, the
overflow bug may be much easier for a SAT solver in a bounded model
checker to detect.  Model checking the code with the added assumption
that {\tt 0 <= m < 3600}, however, the model checker will report the
real bug.  Even though 1 did not appear in the test suite, the
generalization algorithms in the invariant detector generalized the
actual input range to include it.  Extremely large values of {\tt m}
also did not appear during test suite execution, of course, but the
range observed was sufficiently small that invariant detection
proposed that the range was likely restricted.

An obvious objection to this approach is that the invariants from a
test suite will often over-constrain the behavior of many small
components, unless the test suite is very high quality.  We hope that
even highly inaccurate invariants may allow a model checker to find a
real bug.  For example, in the case of {\tt f}, if the test suite
falsely suggests that {\tt 0 <= m < 60} because tests only run for 1
hour, the model checker can still find the real bug.  Our claim is
that our use of dynamic invariants relies on their ability to crudely
``carve away'' the data values that a function is \emph{not} expected
to work rather than on their strict accuracy.  A gross
underapproximation of behavior can still lead to the discovery of real
bugs, while an overappoximation buries a user in false positives.  We
provide suggestive evidence for the possibility that poor invariants
can still be useful by applying our technique to a simple container
class and a radically non-representative test suite.  A more complex
example with hand-generated ``invariants'' shows that even simple
triage can discover useful targets for verification and avoid false
positives using invariants.

In the remainder of the paper, we first discuss key related work in
Section \ref{related}, to place our core idea in context, elaborate
our proof-of-concept example in Section \ref{method}, demonstrate
verification triage in Section \ref{triage}, and discuss the
challenges in bringing the full vision of a ``static analysis-like''
approach to modularity for model checking in Section \ref{discussion}.

\section{Related Work}
\label{related}

To our knowledge, the idea of mining large code bases for arbitrary
sub-components suitable for model checking has not been previously
explored.  The most closely related ideas are efforts to scale model
checking by approximating the \emph{weakest precondition} under which
a sub-component satisfies its correctness properties~\cite{giannakopoulou2012symbolic,alur2005symbolic,cobleigh2006breaking}.
The primary difference is that where these methods aim to learn or
iterate to a ``good'' assumption that is as weak as possible, we
simply want a precondition that removes false positives, even if it is
too strong, and, of course, that such methods have no concept of mining a code base for possible targets for model checking.

\section{Example:  Binary Trees}
\label{method}
\begin{figure}[t]
\begin{verbatim}
public void add(int x) {
    Node current = root;
    if (root == null) {
       br0 = br0 + 1;
       root = new Node(x);
       return;
    }
    while (current.value != x) {
       if (x < current.value) {
            if (current.left == null) {
              br1 = br1 + 1;
              current.left = new Node(x);
            } else {
              br2 = br2 + 1;
              current = current.left;}
            } else {
            if (current.right == null) {
              br3 = br3 + 1;
              current.right = new Node(x);
            } else {
              br4 = br4 + 1;
              current = current.right;
            }
       }
    }
}
\end{verbatim}
\caption{\small Function \texttt{add} after instrumentation for branch count.}
\label{code:add}
\end{figure}

\begin{figure}[t]
\begin{verbatim}
  Random rand = new Rand(20);
  for(i = 0; i < N; i ++){
    BinTree SUT = new BinTree();
    for(int j = 0; j < M; j++){
      int op = rand.nextInt(3);
      int value = rand.nextInt(20);
      switch(op){
         case 0:
           SUT.add(value);
           break;
         case 1:
           SUT.remove(value);
           break;
         case 2:
           SUT.find(value);
           break;
      }
      assert(SUT.RepOK());
    }
  }
\end{verbatim}
\caption{\small A simple random tester for the \texttt{BinTree}.}
\label{fig:BinTreeTester}
\end{figure}

As a simple example of how our approach might work in practice,
assuming that an effective verification task triage can be designed,
we apply our method to a frequently studied binary tree
implementation~\cite{Willemissta2006,Jacoissta2012}.  We use a {\tt
RepOK} function to test the validity of the tree structure, and have
implemented a simple random testing system to test it, shown in Figure
\ref{fig:BinTreeTester}.  We choose to verify the {\tt remove}
function, and instrument the code for the binary tree with
\emph{extended invariants}~\cite{woda2012}, as shown in Figure
\ref{code:add}/footnote{Instrumentation for {\tt add}, not {\tt remove} is
shown as it is easier to follow.} to help us limit the behavior of the
code to valid trees only.  We use Daikon~\cite{daikon} to infer
invariants from all traces generated during random testing.  Extended
invariants introduce history variables that allow Daikon to provide
``invariants'' in the form of code coverage facts as well as more
traditional data invariants, to capture relationships such as, e.g.,
that a loop always executes a number of times equal to twice the value
of a certain input.  Unfortunately, when running the random tester for
5,000 tests, we choose to use tests with a maximum length of only 4
steps, which leads to extended invariants that radically
underapproximate the behavior of the class:
\begin{verbatim}
this.br1 one of { 0, 1, 2 }
this.br3 one of { 0, 1, 2 }
\end{verbatim}

Using these invariants (which restrict the structure to up to two
right children or up to two left children), we generate a simple CBMC harness, knowing (thanks to our triage and invariant examination) that a shallow loop unwinding will suffice, and that no types encountered are likely to make verification difficult, as shown in Figure \ref{code:harness}.

\begin{figure}
\begin{verbatim}
void main()
{  int v1,v2,v3,v4; // symbolic inputs

   /* Using calls for simplicity;
      in practice would encode structure
      from Daikon invariants as a series
      of assumptions. */
   add(v1);
   add(v2);
   add(v3);

   /* Invariant */
   assume(0<=br1 && br1<=2 &&
          0<=br2 && br2<=2);

   remove(v4);
   assert(repOK);
}
\end{verbatim}
\caption{\small CBMC Harness for \texttt{BinTree}.}
\label{code:harness}
\end{figure}

With this extremely restricted harness, CBMC was able to find the bug
in binary tree~\cite{Jacoissta2012}, despite the fact that the shallow
random testing did not discover the bug.

\section{Simple Verification Task Triage}
\label{triage}

Given that using even poor tests to generate potentially incorrect
invariants for model checking can still lead to fast, effective bug
discovery, the core of our modular verification approach becomes
finding functions where this can be applied to large, real-world code
bases.  Figure \ref{code:boyer} shows a function from the source code
for version 1.6 of Mozilla's SpiderMonkey JavaScript engine.  This
function was automatically identified using a very simple 100-line
Python script that crawls through a set of C source files and
identifies functions that:

\begin{itemize}
\item do contain at least some assert statement or potentially crashing memory dereference (i.e., functions that have a ``specification'' even without human intervention) and
\item do not take parameters or declare local variables that cannot be resolved to simple C types (int, short, long, void, and char).
\end{itemize}

This triage approach is both over-restrictive (in many cases functions
taking structures or arrays as input can be handled easily) and not
restrictive enough (it does not filter out functions with recursion or
references to global values of ``bad'' types).  In practice, we expect
triage to require a deeper static analysis that includes more types
and uses summaries of functions called by a function.  Calling a
``bad'' function is not necessarily a problem, because the return
value may be replaced by a value generated by our invariant generator
in many cases.

\begin{figure*}[t]
\begin{verbatim}
jsint
js_BoyerMooreHorspool(const jschar *text, jsint textlen,
                          const jschar *pat, jsint patlen,
                          jsint start)
{
        jsint i, j, k, m;
        uint8 skip[BMH_CHARSET_SIZE];
        jschar c;


        JS_ASSERT(0 < patlen && patlen <= BMH_PATLEN_MAX);
        for (i = 0; i < BMH_CHARSET_SIZE; i++)
            skip[i] = (uint8)patlen;
        m = patlen - 1;
        for (i = 0; i < m; i++) {
            c = pat[i];
            if (c >= BMH_CHARSET_SIZE)
                return BMH_BAD_PATTERN;
            skip[c] = (uint8)(m - i);
        }
        for (k = start + m;
             k < textlen;
             k += ((c = text[k]) >= BMH_CHARSET_SIZE) ? patlen : skip[c]) {
            for (i = k, j = m; ; i--, j--) {
                if (j < 0)
                    return i + 1;
                if (text[i] != pat[j])
                    break;
            }
        }
        return -1;
}
\end{verbatim}
\caption{\small Verification target from Mozilla SpiderMonkey 1.6 source.}
\label{code:boyer}
\end{figure*}

Even using this simplistic version of triage, however, we can obtain
results.  Reading through the roughly 40K lines of non-comment C code
in SpiderMonkey 1.6 would be a time-consuming task for a human, even
ignoring the need to resolve typedefs.  The 11 functions chosen as
possible verification targets by the simple algorithm include some
unsuitable targets (including one case where the textual type analysis
fails to notice a complex compiler state structure parameter).  A few
functions are simple bit-twiddling or offset-computing code that can
be trivially verified without invariants.  The code in Figure
\ref{code:boyer} however is of considerably more interest.  It
implements the Boyer-Moore-Harspool substring-finding algorithm
\cite{BMH}.  After resolving the types to ground C types and choosing
string sizes based on jsfunfuzz tests, we were able to incrementally
increase loop bounds to not only verify memory safety for chosen
strings, but to ``discover'' the worst case complexity of the
algorithm, the point where unwinding assertions held.  When we
introduce an off-by-one error into the code (a case that should not be
found by our testing due to the values chosen for {\tt start}, we
believe), CBMC instantly detects the error.

On the one hand, because the size of strings analyzed is relatively
small, this is not a complete verification of the Boyer-Moore-Horspool
implementation in SpiderMonkey.  On the other hand, in practice for
bug finding, an informal ``small model'' assumption suggests that most
code that works for \emph{all} small inputs works for large inputs, at least
short of integer overflow problems.  Note that the small input size
bounds in our verification are used to also generate small array
sizes, so we can detect memory safety problems with small inputs,
because our memory bounds are also artificially small due to the way
we translate the code into a verification problem.  One problem with
this approach is that when small inputs can lead to larger outputs,
and this invariant is not detected, our approach might result in false
positives.  It remains to be seen how important this is in practice.

That a Boyer-Moore-Horspool implementation is the most interesting
verification target detected by our initial triage system is fitting.
The problem of finding a substring is often discussed (as in the
Wikipedia entry for the algorithm) as the ``needle in a haystack''
problem with the pattern to find called the ``needle.''  Our approach
is essentially the search for model-checkable needles in the haystacks
of large, complex software systems not otherwise amenable to formal
verification without large investments of expert effort.

\section{Core Challenges and Future Work}
\label{discussion}
We speculate that while model checking cannot apply to as many
functions as static analysis, our method may be of considerable value
for bug detection because of the precision and complete exploration of
model checking.  However, the critical question is whether dynamic
invariants can actually produce accurate enough preconditions to
reduce the false positive rate to a manageable level.  Determining
whether dynamic invariants over even poor test suites can still find
bugs for realistic systems requires the development of a tool for
automatic verification task triage.  Our initial efforts suggest that
this task is complex, though not infeasible.  In particular, our early
efforts to triage C programs have shown that the first steps are
identifying code patterns that frustrate a particular model checker,
and this requires a complete analysis of the accessed data types and
call graphs from each function for ``bad'' patterns, like complex
recursive data structures or system calls.  We have
identified some likely candidate functions in the Mozilla JavaScript
engine and SQLite, and plan to use these modest sized code bases to
tune a method for handling larger programs with more complex build
environments.  The next steps are continued development of the triage
tools and experimentation with actual invariants from the test suites
for our subject programs.

\bibliographystyle{plain}
\bibliography{ref.bib}

\end{document}